\documentstyle[headers,epsf]{article}
\begin{document}
\twocolumn[\hsize\textwidth\columnwidth\hsize\csname @twocolumnfalse\endcsname
\begin{center} 
{\large NEAR-FIELD SCANNING MICROWAVE MICROSCOPY: MEASURING LOCAL MICROWAVE PROPERTIES AND ELECTRIC FIELD DISTRIBUTIONS}
\end{center}    
\begin{center}
B. J. Feenstra, C. P. Vlahacos, Ashfaq S. Thanawalla, D. E. Steinhauer, S. K. Dutta, 
F. C. Wellstood and Steven M. Anlage
\end{center}
\begin{center}        
	{\it Center for Superconductivity Research, Department of Physics,\\
	University of Maryland, College Park, MD 20742-4111, USA}
\end{center}
\vskip2pc]
\section*{Abstract}
We describe the near-field scanning microwave microscopy of microwave 
devices on a length scale much smaller than the wavelength used for imaging.
Our microscope can be operated in two possible configurations, allowing 
a quantitative study of either material properties or local electric fields.

\section*{Introduction}

In recent years, several novel methods of scanning probe microscopy have been 
developed. Generally speaking one can divide these methods into two main approaches,
driven by distinct motivations.
In the first approach the main goal is to study material properties, preferably on the 
smallest possible length scale.
Examples of such techniques are atomic force microscopy (AFM) and scanning tunneling microscopy (STM) \cite{art:Balk94,art:Amato97}, which have achieved atomic scale spatial resolution.
The second objective is the investigation of fields, both electric and magnetic, emitted 
by operating devices. An example of such a technique is scanning microwave
microscopy\cite{art:Budka96,art:Dutta98}.
For devices operating in the microwave range, a scanning technique
at microwave frequencies will be useful for diagnosing circuit problems. 
Despite the relatively long wavelength, 
of the order of mm's to cm's, spatial resolution below 1 $\mu$m can be achieved 
by operating in the so-called near-field limit\cite{art:Bryant65,art:Ash72}.
In this paper we will illustrate the versatility of near-field scanning microwave microscopy, by showing results on local material properties as well as on electromagnetic field imaging in close proximity to operating devices.

\section*{Experimental Setup}

\begin{figure}[htb]
\begin{center}
\leavevmode
\epsfxsize=7.5cm
\epsffile{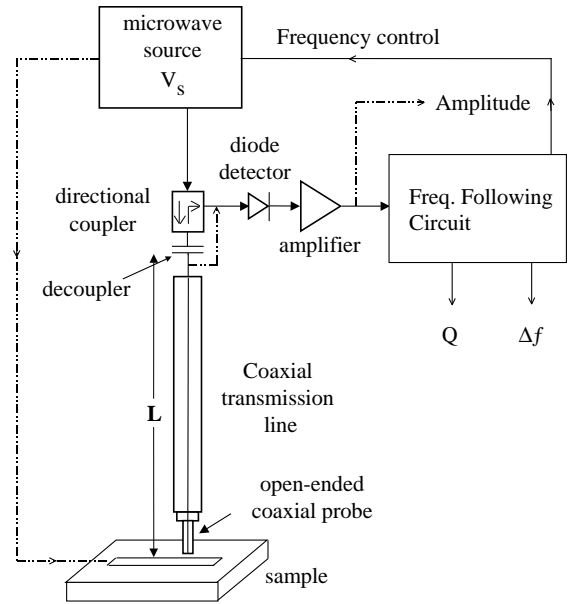}
\end{center}
\caption{{\it Experimental setup, showing the reflecting mode (solid lines) 
and the receiving mode (dash-dotted lines).}}
\label{setup}
\end{figure}

Fig.\ \ref{setup} shows a schematic of our experimental setup. 
The microwave microscope can be operated in two distinct configurations:
what we call the reflecting mode (solid lines in Fig. \ref{setup}) and the 
receiving mode (dashed-dotted lines). 

\subsection*{Reflecting Mode}

In the reflecting mode, a microwave signal is fed through a coaxial transmission line 
which has been capacitively coupled to the source and ends in a 
rigid open-ended coaxial probe. 
This creates a resonant circuit, in which the resonant frequency $f_0$ and 
the quality factor $Q$ are modified by the sample which is 
in close proximity to the open end of the probe. 
Using a lock-in based frequency-following circuit\cite{art:Steinhauer97} 
we measure the shift of the resonance $\Delta f$ as the sample is scanned 
under the probe. We can also measure the strength of the signal at twice 
the lock-in modulation frequency.
The latter is related to the curvature of the resonance peak and is
a measure of the $Q$ of the system\cite{art:Steinhauer98}. 
By measuring both quantities simultaneously we obtain information 
about the real and the imaginary part of the complex dielectric function, 
$\epsilon$, of the sample.

\subsection*{Receiving Mode}

A slightly modified configuration (dash-dotted lines in Fig.\ \ref{setup}) 
can be used to measure electric fields 
emitted by operating microwave devices.
In this case, microwave power is fed directly to the device and the coaxial probe 
senses the electric fields from the sample. 
The microwave power measured by the probe is fed to the diode detector directly, bypassing the decoupler, directional coupler and the resonant transmission line. 
The measured amplitude can be converted to electric field, using additional  
information about the setup\cite{art:Dutta98}.

\section*{Material Properties}

The data analysis can be complicated considerably when, 
in addition to the variations in material properties, the sample exhibits topographic 
features\cite{art:Knoll97}. 
Several methods have been used to separate out the topography. 
The simplest approach is to scan the same region
twice, initially recording the topography\cite{art:Strausser94}.
Another solution is to measure the probe-sample separation independently, and 
use this information in a feedback loop to maintain the probe at a constant 
height\cite{art:Hsu95}. 

\begin{figure}[h!]
\begin{center}
\leavevmode
\epsfxsize=6.6cm
\epsffile{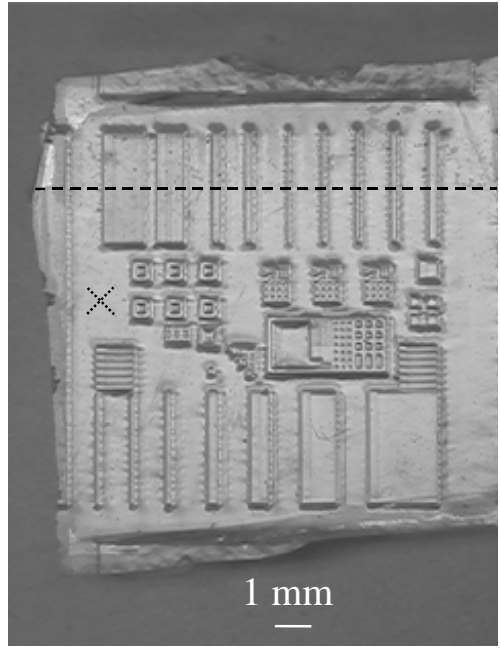}
\end{center}
\caption{{\it Photograph of a glass chip which has a variety of structures etched 
into the surface. The cross indicates the position at which the calibration curve in 
Fig.\ \protect{\ref{dielcali}} was measured.}}
\label{dielsamp}
\end{figure}

\begin{figure}[h!]
\begin{center}
\leavevmode
\epsfxsize=8cm
\epsffile{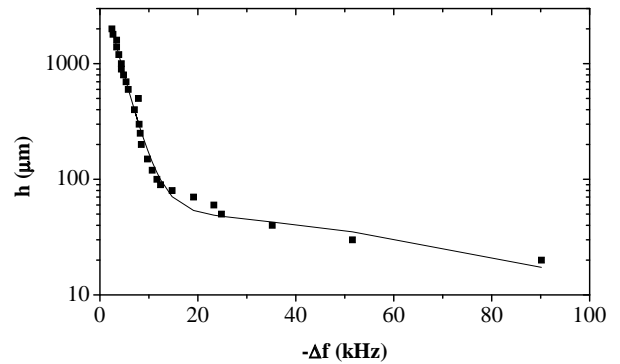}
\end{center}
\caption{{\it Calibration curve correlating frequency shift $\Delta f$ and height $h$,
measured at 9.6 GHz using a 100 $\mu m$ probe.}}
\label{dielcali}
\end{figure}

Our system is especially sensitive to topographic features when we measure 
the frequency shift $\Delta f$. 
Figure \ref{dielsamp} shows a glass chip in which various 
topographic features were etched. 
By measuring the height dependence of the frequency shift at one 
particular spot on the sample (Fig.\ \ref{dielcali}), we calibrate the 
relation between $\Delta f$ and the height $h$. 
Next, we fit the response 
to an empirical function, shown as the solid line. 
We can then use this 
function to determine the absolute height of the probe above the 
entire sample\cite{art:Vlahacos98}. 

\begin{figure}[htb]
\begin{center}
\vspace*{.5cm}
\leavevmode
\epsfxsize=8cm
\epsffile{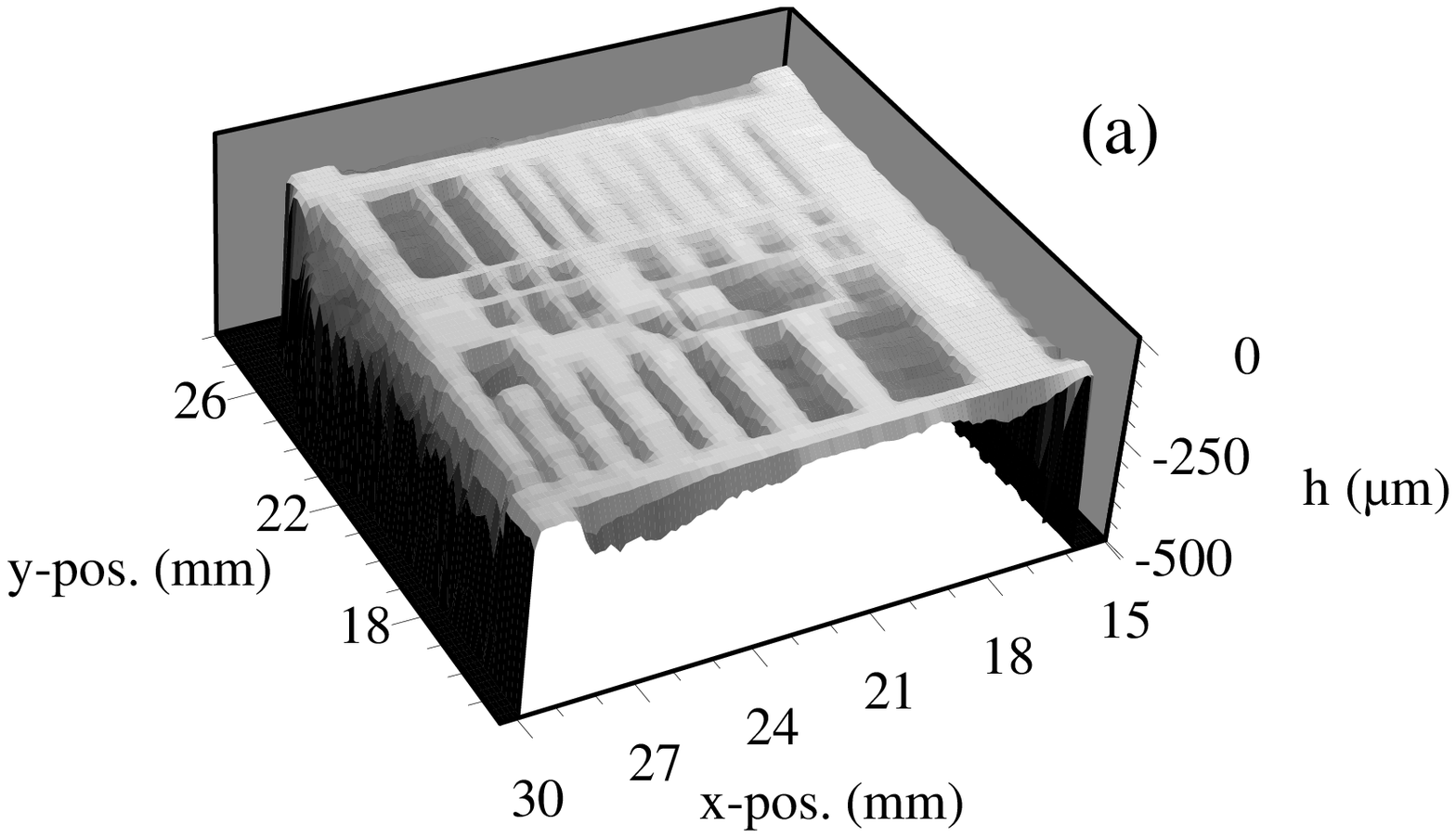}
\end{center}
\begin{center}
\leavevmode
\epsfxsize=8cm
\epsffile{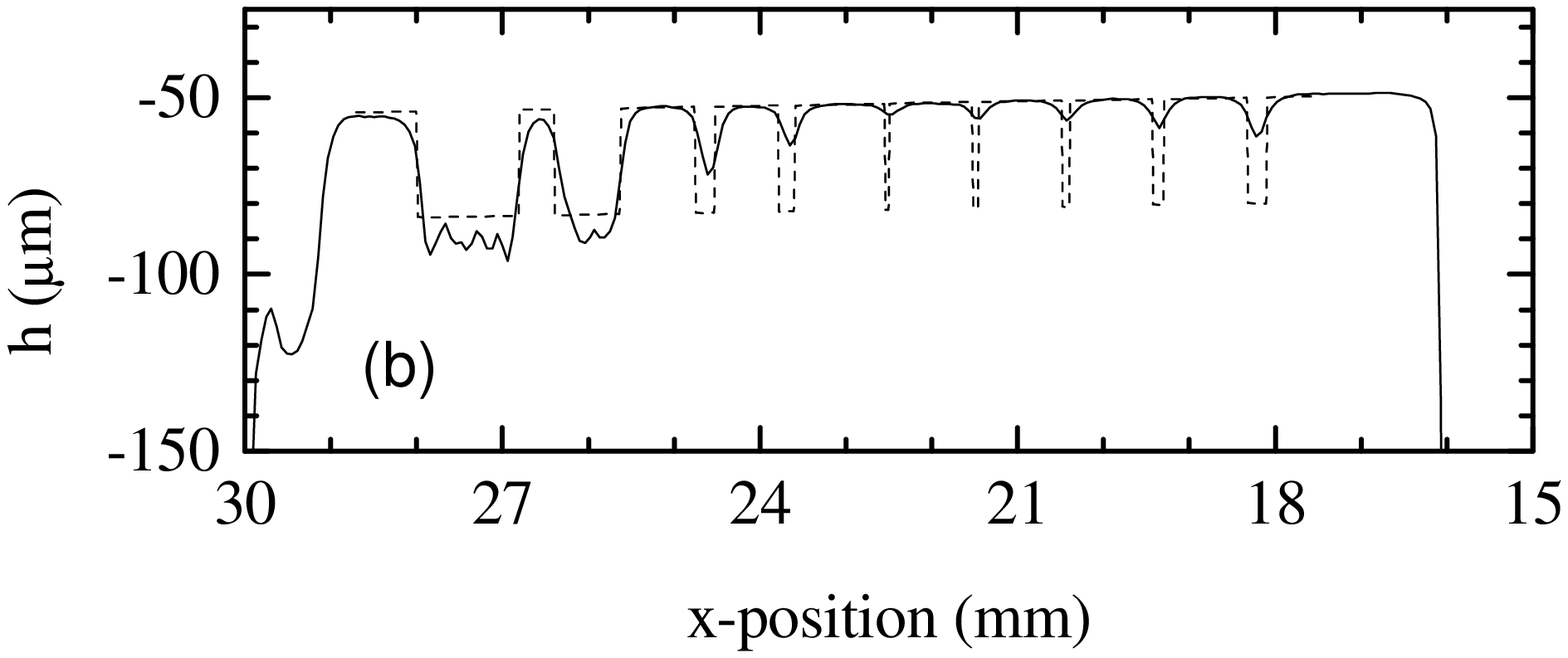}
\end{center}
\caption{{\it (a): Topographic image of the glass chip, as measured with the microwave
microscope at 9.6 GHz. (b): Line cut, taken along the dashed line in Fig.\ \protect{\ref{dielsamp}}, showing the depth profile in more detail.}}
\label{dielimag}
\end{figure}

Figure \ref{dielimag}a shows such a topographic image taken 
using a probe with an inner conductor diameter of 100 $\mu m$
at 50 $\mu m$ above the highest point on the sample.
We are able to discern height changes as small as 55 nm, this figure being limited 
by the frequency stability of the microwave source.
A line cut, taken along the dashed line in Figure \ref{dielsamp}, reveals more details (see Fig.\ \ref{dielimag}b). The dashed line shows the width of the slits, all of which are
30 $\mu m$ deep.
A fair agreement is obtained for most
of the line-cut, although it can be seen that for features comparable to 
or smaller than the probe size, the height measured with the microwave microscope
suffers from averaging effects, yielding results which are too shallow.
The effective spatial resolution is $d \sim 100 \mu m \approx \lambda/300$.

\section*{Electric Field Imaging}

An important potential application of microwave microscopy is for the diagnosis of
problems in integrated circuits. Since many circuits operate at microwave 
frequencies, it is essential to have a technique that searches for faults at 
the operating frequency. In addition, a non-destructive technique is desired.
Near-field microwave microscopy offers the possibility of scanning the coaxial 
probe in close proximity to an operating device, thereby sensing the 
electric field component normal to the face of the probe\cite{art:Dutta98}.

\begin{figure}[htb]
\begin{center}
\leavevmode
\epsfxsize=7.5cm
\epsffile{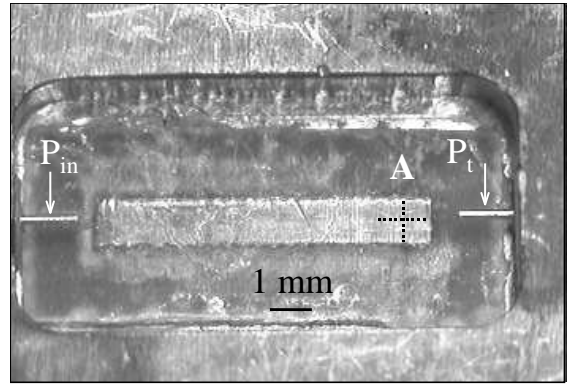}
\end{center}
\caption{{\it Photograph of Cu-microstrip resonator. Also indicated are the 
position at which the power is capacitively coupled into the resonator (P$_{in}$) and 
the position at which the frequency response was measured (A). The chip is enclosed
in a Cu-package.}}
\label{Cuphoto}
\end{figure}

\begin{figure}[htb]
\begin{center}
\leavevmode
\epsfxsize=7.5cm
\epsffile{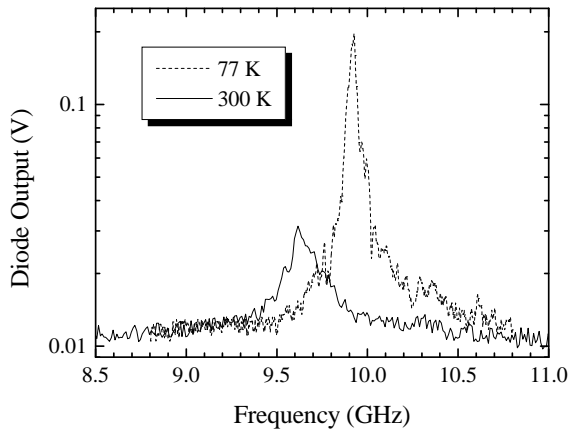}
\end{center}
\caption{{\it Frequency response of the Cu-microstrip, at 300 K (solid line) and 
77 K (dashed line).}}
\label{Cufre}
\end{figure}

Much of our research is devoted to the study of superconducting devices. For this reason, we have developed a microscope which works at temperatures down to 4.2 K\cite{art:Thanawalla98}.
In Fig.\ \ref{Cuphoto} we show a Cu-microstrip resonator. 
The groundplane of the microstrip is also made of copper, 
and the microstrip dimensions are 8$\,\times\,1$ mm. Fixing a 200 $\mu m$ 
diameter probe at 1 mm above position A, and coupling the microwave power in at $P_{in}$, 
we measured the signal picked up by the probe as a function of frequency.
The response at 300 K and 77 K is shown in Fig.\ \ref{Cufre}. 
Upon lowering the temperature, both the signal picked up by the probe and the $Q$ 
of the resonator are enhanced considerably.

From Fig.\ \ref{Cufre}, we also see that the resonant frequency of the microstrip increases 
at low temperatures. 
One possible cause is that the decreased inductance (reduced skin depth) 
at lower temperatures will shift the resonant frequency upward.
This is consistent with the reduced losses at 77 K, clearly visible in the enhanced 
quality factor of the resonance.
Assuming that the enhanced $Q$ can be associated
entirely with the Cu-microstrip, we can calculate how large the corresponding frequency 
shift will be. For $Q(77)/Q(300)\!=\!4$ one expects a change in frequency of 
approximately  0.1 \%, giving $\Delta f\!\approx\!10$ MHz. Since 
the measured frequency shift is much larger ($\approx\!300$ MHz), 
it is clear that this effect plays only a minor role.
Secondly, thermal contraction will also increase the resonant frequency. Assuming 
a thermal contraction coefficient $\alpha\!=\!\Delta L/L\!=\!10^{-5}\!/K$, the 
relative change of the length of the resonator would be approximately 
$2\!\times\!10^{-3} $, implying a similar change in the resonant frequency.
Finally, the presence of the probe close to the resonator imposes a perturbation on 
the latter. Since this is an electric field perturbation, the resonant frequency will
tend to shift down. Part of the shift may hence be due to the fact that the coupling 
between the probe and the Cu-resonator is less at low temperatures.

\begin{figure}[h!]
\begin{center}
\leavevmode
\epsfxsize=7.5cm
\epsffile{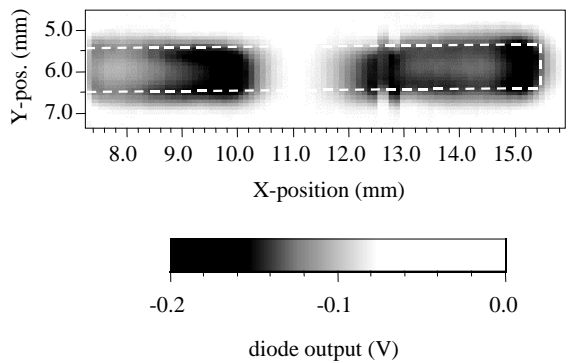}
\end{center}
\caption{{\it Electric field pattern above a 77 K Cu-microstrip line, operating at its
fundamental resonant frequency, 9.90 GHz.
The edges of the stripline are indicated by the dashed lines.}}
\label{Cuxyscan}
\end{figure}

To image the sample, we fixed the frequency at the resonance peak
and scanned the sample underneath the coaxial probe, 
measuring the electric field as a function of position. 
The resulting xy-scan at 77 K can be seen in Fig.\ \ref{Cuxyscan};
the probe was kept at a constant distance of 350 $\mu m$ above the sample. 
We see that the electric field peaks at the ends of the stripline and vanishes in 
the middle, as expected for the fundamental voltage mode. 
Close to the power pin (near x = 7 mm) the perturbing effect of the probe 
is more serious than in the rest of the image, causing
the amplitude to decrease slightly. 

\section*{Conclusions}

We have demonstrated the ability of near-field scanning microscopy to 
provide quantitative information on material properties and 
electric field distributions at microwave frequencies.
We are able to study microwave 
properties at a length scale as small as $\lambda/300$, operating at 
temperatures varying from 4.2 to 300 K.\\
\\
\indent
{\it Acknowledgements}
This work has been supported by NSF-MRSEC grant No. DMR-9632521,
NSF grant No. ECS-9632811 and the Center for Superconductivity Research.

\end{document}